%% file: main.tex
\documentclass[10pt,conference]{IEEEtran}

\usepackage{cite}
\usepackage{amsmath,amssymb,amsfonts}
\usepackage{graphicx}
\usepackage{booktabs}
\usepackage{xcolor}
\usepackage{subcaption}
\usepackage{url}
\usepackage{braket}
\usepackage[hidelinks]{hyperref}
\input{temp-definitions}
\usepackage[thinc]{esdiff}
\title{Variational Quantum Algorithms for Hyperelasticity: Incorporating Nonlinear Constitutive Behavior}

\author{
\IEEEauthorblockN{Uditnarayan Kouskiya}
\IEEEauthorblockA{
Department of Civil and Environmental Engineering\\
Vanderbilt University\\
1001 19th Ave S, Nashville, TN 37212\\
uditnarayan.kouskiya@vanderbilt.edu
}
\and
\IEEEauthorblockN{Caglar Oskay}
\IEEEauthorblockA{
Department of Civil and Environmental Engineering\\
Department of Mechanical Engineering\\
VU Station B\#351831, 2301 \\ Vanderbilt Place, Nashville, TN 37235\\ caglar.oskay@vanderbilt.edu
}
}

\begin{document}

\maketitle

\begin{abstract}
This paper extends a recently proposed Variational Quantum Algorithm (VQA) framework for nonlinear elasticity to a broader class of constitutive nonlinearities involving rational powers of the stretch. One-dimensional incompressible Ogden and Mooney–Rivlin models are employed as representative examples to demonstrate the proposed methodology. Nonlinear constitutive terms are transformed into forms compatible with the available quantum algorithmic primitives through the introduction of auxiliary variables and penalty constraints, yielding approximate solutions via a VQA. An iterative correction strategy based on a sequence of VQAs is then introduced to improve solution accuracy. A Numerical example demonstrates the proposed approach. 
\end{abstract}

\section{Introduction}
Recent advances in quantum computing have motivated the development of quantum algorithms for differential equations. However, two major challenges remain: many algorithms with strong theoretical guarantees require fault-tolerant hardware, impractical on near-term devices, and the linear structure of quantum state evolution complicates the treatment of nonlinear differential equations. To address these challenges, we previously proposed a Variational Quantum Algorithm (VQA) framework for elasticity problems governed by hyperelastic constitutive models, building on the VQA formulations previously proposed for linear problems (e.g.,~\cite{Arora:2025a}). The framework reformulates the elastostatic potential energy as a cost function for the VQA and was demonstrated using a one-dimensional compressible Neo-Hookean model, where the logarithmic constitutive nonlinearity was treated through a polynomial approximation~\cite{KO24}. The present work extends the framework to constitutive models involving fractional and inverse powers of the stretch, with the incompressible Ogden and Mooney–Rivlin models serving as representative examples, and introduces an iterative correction strategy to further improve solution accuracy.

The remainder of this paper is organized as follows: Section \ref{sec:VQA_outline} briefly reviews the framework. Section \ref{sec:methods} presents the proposed approximation and iterative correction strategy, while Section \ref{sec:examples} demonstrates its application to incompressible Ogden and Mooney–Rivlin models. Concluding remarks are provided in Section \ref{sec:conclusion}.

\section{VQA Framework for 1D hyperelasticity} \label{sec:VQA_outline}
\subsection{Mathematical Model} \label{sec:math_model}
The nonlinear elastostatic boundary-value problem in the reference configuration $\bbOmega_{0}\subset\mathbb{R}^{3} (\bbX = (X,Y,Z))$ is given by
\begin{equation*}
\label{eq:governing_1}
\nabla_{\bbX}\cdot\bbP(\bbF(\bbX)) + \bbB(\bbX) = \mathbf{0}
\quad \text{for }\bbX \in \bbOmega_{0},
\end{equation*}
where $\bbP$, $\bbF$ and $\bbB$ are the first Piola-Kirchhoff stress tensor, deformation gradient and the body force per unit reference volume, respectively. For hyperelastic materials, the first Piola-Kirchhoff stress can be derived from a strain energy density function $W(\bbF)$ via $\bbP = \partial W/\partial\bbF$. 

We consider the problem with reference configuration to be a prismatic body of constant cross-sectional area $A$, with all fields depending only on the axial coordinate $X$. The lateral surface is traction-free, while displacement and/or traction boundary conditions are prescribed on the end faces, $\Gamma_D$ and $\Gamma_N$, respectively. The force balance along the axial direction is given by 
\begin{subequations}\begin{gather}
\label{eq:1D_force_balance}
\diff{}{X}\big(P(\lambda(X))\big)+B(X)=0
\quad\text{in }\Omega_0 = (0,L); \\ 
u=\bar u \quad\text{on }\Gamma_D,
\qquad
P=\bar P \quad\text{on }\Gamma_N.
\end{gather}
\end{subequations}
For incompressible materials,
$$
J=\det(\mathbf F)=1,
\qquad
\lambda_2=\lambda_3=\lambda^{-1/2},
$$
where $\lambda=\lambda_1$ denotes the axial stretch. Consequently, the three-dimensional strain-energy density reduces to a one-dimensional function $W(\lambda)$. The corresponding potential energy functional is
\begin{equation}
\mathcal{E}[u]
=
\int_{\Omega_0} [ W(\lambda)
- B(X)\,u(X)]\,\mathrm dX
-
\sum_{X\in\Gamma_N}\bar P\,u(X),
\label{eq:mardsen_1d}
\end{equation}
where we will refer to the total contribution by loading terms using $\mathcal{E}_B[u]$. The critical point of \eqref{eq:mardsen_1d} can be established to be a weak solution of the governing equation \eqref{eq:1D_force_balance}.

\begin{figure}[t]
    \centering
    \includegraphics[width=0.99\columnwidth]{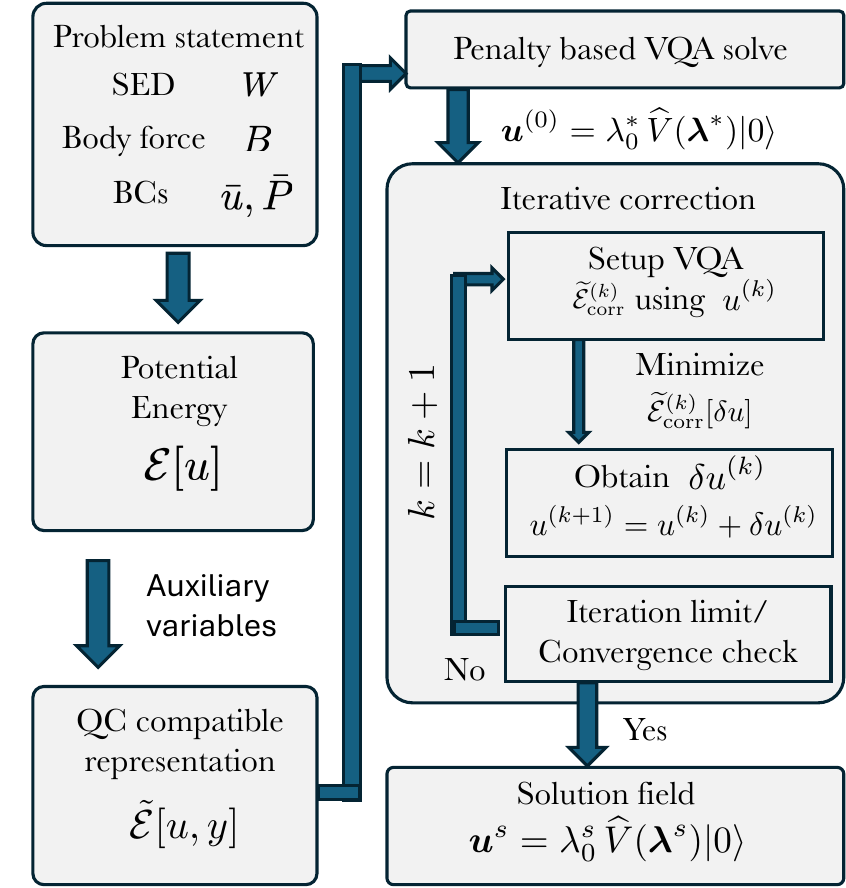}
    \caption{Overview of the proposed approach. $u^s$ represents the final solution field.}
    \label{fig:framework}
\end{figure}

\begin{figure}[t]
    \centering
    \includegraphics[width=0.8\columnwidth]{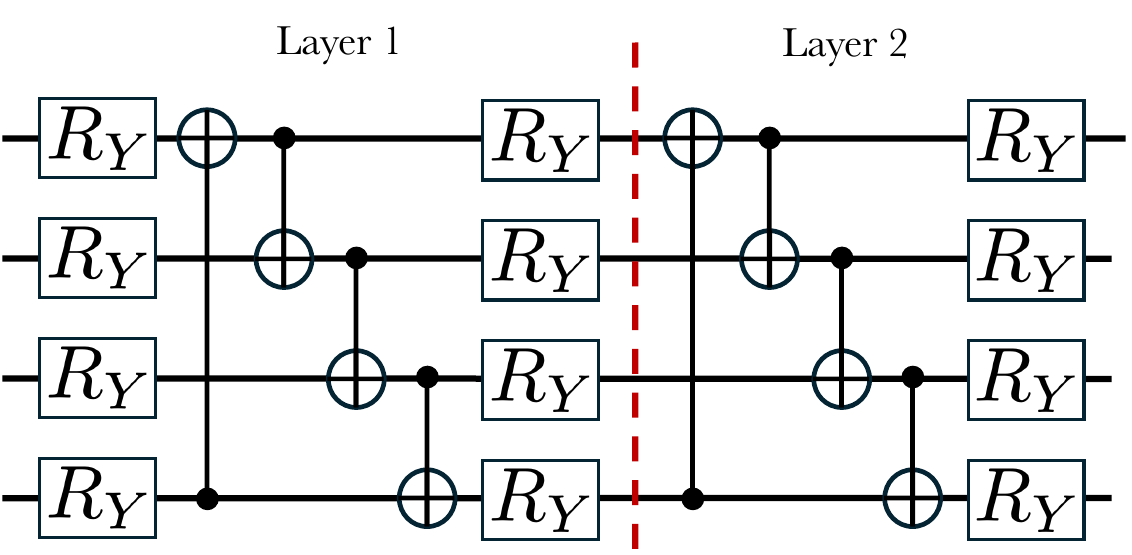}
    \caption{Parameterized quantum circuit (ansatz) employed in the present work by Sim et al.~\cite{sim_et_al}.}
    \label{fig:ansatz}
\end{figure}

\begin{figure}[t]
    \centering
    \includegraphics[width=0.7\columnwidth]{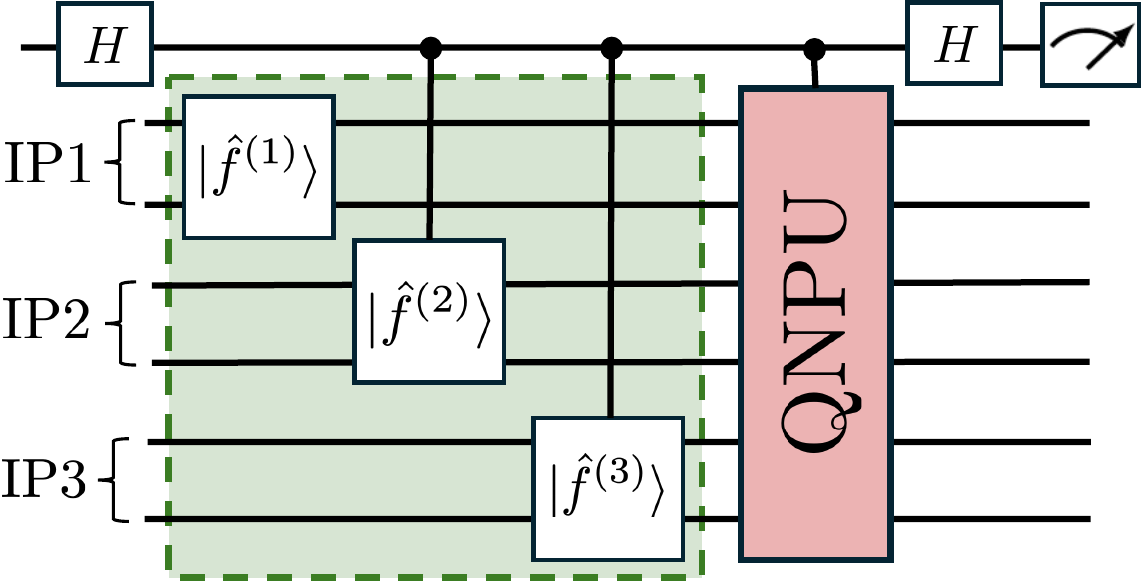}
    \caption{Quantum nonlinear processing unit (QNPU) framework used to evaluate polynomial contributions appearing in the cost function.}
    \label{fig:qnpu}
\end{figure}

\subsection{Variational Quantum Algorithm}
Figure \ref{fig:framework} summarizes the framework, and the details can be found in Ref.~\cite{KO24}. The proposed approach reformulates the discretized potential energy \eqref{eq:mardsen_1d} as the cost function of a variational quantum algorithm. Following spatial discretization, the displacement DoF vector is encoded into a quantum state according to
\begin{equation*}
\bbu
=
\|\bbu\|\,|\hat{u}\rangle
=
\theta_0\,\widehat V(\boldsymbol{\theta})|0\rangle,
\end{equation*}
where $\widehat V(\boldsymbol{\theta})$ denotes a parameterized quantum circuit (ansatz), $\boldsymbol{\theta}$ is the set of variational parameters, and $\theta_0$ is a normalization factor. 
For all numerical experiments, we employ the ansatz proposed by Sim et al.~\cite{sim_et_al} with a circuit depth of $d=4$, as illustrated in Figure~\ref{fig:ansatz}. 
The quantities $(\theta_0,\boldsymbol{\theta})$ are collectively referred to as the control parameters and determine the discrete displacement field. The potential energy is thereby represented as the quantum cost function
\begin{equation*}
\mathcal C(\theta_0,\boldsymbol{\theta})
:=
\mathcal E
\big(
\bbu(\theta_0,\boldsymbol{\theta})
\big).
\end{equation*}
The cost function is evaluated through a small set of quantum circuits that estimate the required expectation values, which are subsequently combined classically to recover the total energy of the system.

A classical gradient-based optimizer is employed to iteratively update the control parameters in search of a critical point of the cost function. The gradients $(\partial\mathcal C/\partial\theta_i)$ are computed using finite differences, with each cost evaluation requiring the same set of quantum circuits used to evaluate $\mathcal C$. The optimization is terminated using a cost-based stopping criterion $\mathcal F_{\mathrm{opt}}$. The resulting optimal parameters $(\theta_0^*,\boldsymbol{\theta}^*)$ define the discrete displacement field
\begin{equation*}
\bbu^*
=
\theta_0^*\,\widehat V(\boldsymbol{\theta}^*)|0\rangle,
\end{equation*}
which corresponds to a critical point of the discretized potential energy and is therefore taken as the solution. For the one-dimensional hyperelastic models considered in this work, convexity of the strain-energy density $W$ (equivalently, a positive tangent stiffness) guarantees a unique minimizer \cite{marsden1994mathematical}. Consequently, the cost function is treated as a minimization problem throughout this study. Details of the quantum primitive circuits and their synthesis into a complete cost-function evaluation framework are provided in Section~2 of Ref.~\cite{KO24}. Since the underlying quantum primitives remain unchanged, the same construction principles can be employed to realize the cost functions arising from the constitutive models considered in the present study.

\section{Implementing Nonlinear Constitutive Laws under the VQA Framework}
\label{sec:methods}

Arbitrary nonlinearities appearing in the strain-energy density $W$ cannot, in general, be directly represented using the currently available shallow-circuit VQA framework. However, Quantum Nonlinear Processing Units (QNPUs) (Figure~\ref{fig:qnpu}), introduced by Lubasch et al.~\cite{lubasch_zoo}, enable the evaluation of polynomial operations of the form
\begin{equation}
\label{eq:QNPU_operation}
E
=
\hat{f}^{(1)\,\top}
\left(
\odot_{j=1}^{r}
O_j \hat{f}^{(j)}
\right),
\end{equation}
where $O_j$ denotes a linear operator acting on a unit-norm vector $\hat f^{(j)}\in\mathbb R^{N_q}$, and $\odot$ represents the element-wise (Hadamard) product. Here, $N_q$ denotes the Hilbert space dimension generated by an $n$-qubit system.

For the present application, this capability allows the evaluation of integral quantities of the form
\begin{equation}
\label{eq:reduced_form}
\int_{\Omega_0}
\prod_{i=1}^{r}
O_i f_i
\, dX,
\end{equation}
where $O_i$ are linear differential operators acting on fields $f_i(X)$. Following spatial discretization, such integrals reduce to algebraic expressions of the form \eqref{eq:QNPU_operation} involving the corresponding DoF vectors. Consequently, the control parameters defining the quantum state may be supplied directly to the quantum circuits, whose expectation values are then combined classically to evaluate the cost function.

The proposed solution strategy consists of two stages. First, the potential energy $\mathcal{E}$ is represented in a form compatible with \eqref{eq:reduced_form}, potentially requiring an approximation of $W$, and subsequently minimized using a VQA to obtain an approximate displacement field. Second, a sequence of correction VQAs is employed to iteratively improve the resulting solution.

\subsection{Polynomial Approximation using Penalty Methods}

We consider strain-energy densities of the form
\begin{equation}
W(\lambda)
\approx
\sum_{i=0}^{N_{t_n}}
c_i\,\lambda^{p_i},
\label{eq:general_nonlin}
\end{equation}
where $c_i$ are material-dependent coefficients and $p_i\in\mathbb{Q}$. For constitutive laws that do not naturally admit the representation \eqref{eq:general_nonlin}, suitable approximations may be employed. For example, the logarithmic nonlinearity arising in the compressible Neo-Hookean model can be represented through Taylor- or inverse hyperbolic tangent-based expansions, as discussed in Ref.~\cite{KO24}.

To eliminate fractional powers appearing in \eqref{eq:general_nonlin}, auxiliary variables are introduced such that the resulting constitutive contributions become polynomial in an augmented set of variables. For
\begin{equation*}
p_i=\frac{a_i}{b_i},
\qquad
a_i,b_i\in\mathbb Z,
\qquad
b_i\neq 0,
\end{equation*}
we introduce an auxiliary variable $y$ to enforce 
$y=\lambda^{a_i/b_i}.$
The corresponding constitutive contribution can then be expressed as a polynomial in $y$. Consequently, the total potential energy is modified as
\begin{equation*}
\widetilde{\mathcal E}[u,y]
=
\mathcal E[u,y]
+
\frac{\beta}{2}
\int_{\Omega_0}
\left(
y^{b_i}-\lambda^{a_i}
\right)^2
\,dX,
\end{equation*}
where $\mathcal E[u,y]$ is obtained from $\mathcal E[u]$ by replacing $\lambda^{p_i}$ with $y$ and $\beta$ represents the penalty parameter. For exponents $a_i$ and $b_i$ of opposite sign, the penalty term may equivalently be written as
\begin{equation*}
\frac{\beta}{2}
\int_{\Omega_0}
\left(
1-y^{b_i}\lambda^{a_i}
\right)^2
\,dX,
\end{equation*}
thereby eliminating negative powers from the constraint equation. Additional control parameters $(\gamma_0,\boldsymbol{\gamma})$ to govern the auxiliary variable $y$ will be required and the optimization problem will get updated to:
\begin{equation*}
(\lambda_0^*,\gamma_0^*,\boldsymbol{\lambda}^*,\boldsymbol{\gamma}^*)
=
\arg\min_{\lambda_0,\gamma_0,\boldsymbol{\lambda},\boldsymbol{\gamma}}
\mathcal{C}
\!\left(
\lambda_0,\gamma_0,
\boldsymbol{\lambda},
\boldsymbol{\gamma}
\right).
\end{equation*}

\subsection{Iterative Correction}
The introduction of the polynomial representation \eqref{eq:general_nonlin} modifies the original potential energy functional $\mathcal E$, thereby introducing an approximation whose accuracy depends on the quality of the underlying representation. Even for strain-energy densities that naturally admit the form \eqref{eq:general_nonlin}, the penalty formulation introduces an additional source of error. Small penalty parameters may lead to poor enforcement of the auxiliary constraints, whereas excessively large values can result in ill-conditioning, as is well known in the optimization literature~\cite{NocedalWright}.

Let $\tilde u$ denote an approximate solution and consider a correction $\delta u$ such that
$$
u=\tilde u+\delta u,
\qquad
\tilde\lambda = 1+\tilde u'.
$$
Assuming $\delta u'$ is sufficiently small, a Taylor expansion of the strain-energy density about $\tilde\lambda$ yields
\begin{equation}
W_T(\lambda(X))
=
W(\tilde\lambda(X))
+
\sum_{i=1}^{N_t}
A_i(\delta u')^i,
\label{eq:iterative_correction}
\end{equation}
%
%

where $A_i(X)$ represent the derivatives of $W$ at $X$ and  $N_t$ denotes the truncation order of the expansion.

Neglecting terms independent of $\delta u$, the correction is obtained as the minimizer of
\begin{equation*}
\widetilde{\mathcal E}_{\mathrm{corr}}[\delta u]
=
\int_{\Omega_0}
\sum_{i=1}^{N_t}
A_i(\delta u')^i\,dX
+
\mathcal{E}_B[\delta u]
\end{equation*}

The coefficient fields $A_i(X)$ are completely determined by the current approximation $\tilde u$ and are therefore treated as known fields over the domain $\Omega_0$ during the correction step. The functional $\widetilde{\mathcal E}_{\mathrm{corr}}$ is minimized using a VQA, yielding an approximation to the correction field $\delta u$. Repeating this procedure through the update
$$
u^{(k+1)}
=
u^{(k)}
+
\delta u^{(k)},
$$
generates a sequence of corrections that progressively improves the solution.

The effectiveness of this procedure is not guaranteed for arbitrary corrections, particularly when $\delta u'$ becomes large. Furthermore, its success depends on the coefficient fields $A_i$, which are determined by the constitutive model and the field $u^{(k)}$ under consideration. For the examples studied herein, the method performs well when initialized from an approximate solution that is sufficiently close to the true solution.

\section{Examples} \label{sec:examples}
Following the incompressible formulation presented in Section~\ref{sec:math_model}, we consider the single-term Ogden model \cite{Jog2015}
\begin{equation*}
W(\lambda)
=
\frac{2\mu}{\alpha^2}
\left(
\lambda^{\alpha}
+
2\lambda^{-\alpha/2}
-3
\right),
\end{equation*}
where $\mu$ is the shear modulus and $\alpha\in\mathbb{R}$. Similarly, the one-dimensional Mooney--Rivlin (MR) model \cite{Jog2015} corresponding to the present assumptions is given by
\begin{equation*}
W(\lambda)
=
C_1
\left(
\lambda^2
+
2\lambda^{-1}
-3
\right)
+
C_2
\left(
2\lambda
+
\lambda^{-2}
-3
\right).
\end{equation*}

\begin{figure}[t]
    \centering

    \begin{subfigure}{\columnwidth}
        \centering        \includegraphics[width=\linewidth]{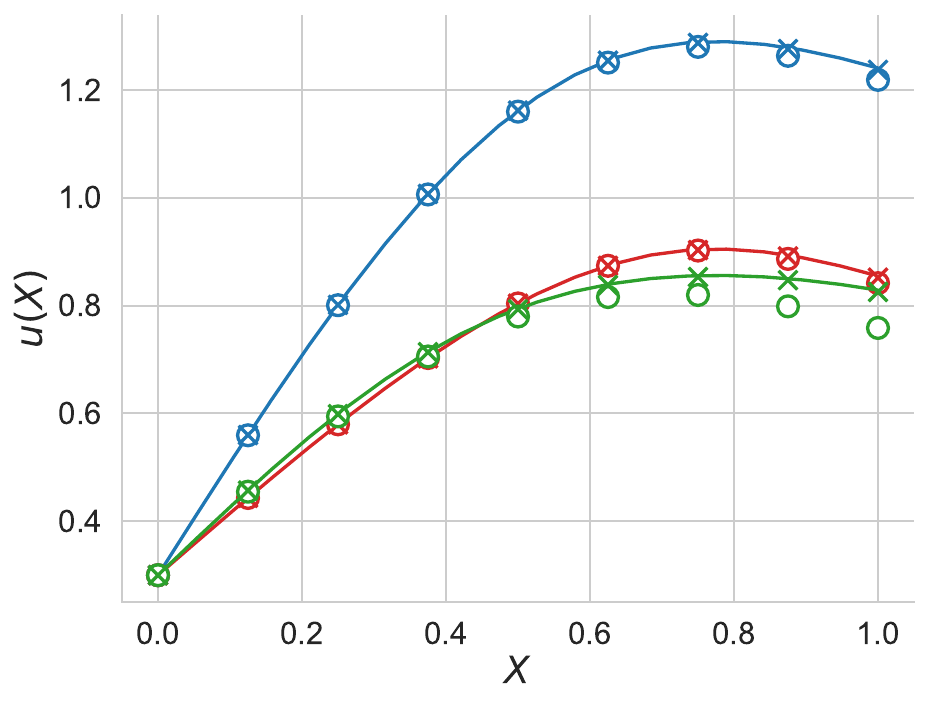}
        \caption{}
        \label{fig:comparison_models_plot}
    \end{subfigure}

    \vspace{0.5em}

    \begin{subfigure}{1\columnwidth}
    \centering
    \includegraphics[width=\linewidth]{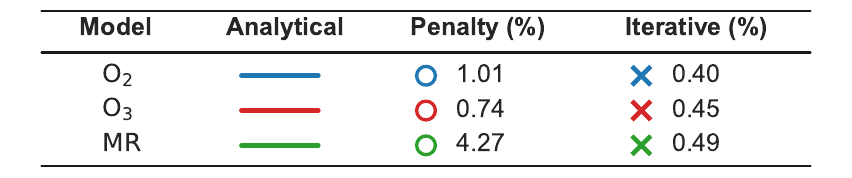}
    \caption{}
    \label{fig:comparison_models_table}
\end{subfigure}

    \caption{
    Comparison of analytical (dotted), penalty-based, and iteratively corrected quantum solutions (mm) for the O$_2$, O$_3$, and MR models.
    (a) Displacement fields from the three approaches.
    (b) Relative $L^2$ errors (\%) vs.\ analytical solution.
    }
\label{fig:comparison_models}\vspace{-0.5cm}
\end{figure}

We consider $\alpha=2$ and $3$ (denoted by $\mathrm{O}_2$ and $\mathrm{O}_3$, respectively) together with an MR model having $C_1=C_2=0.5$. The case $\alpha=2$ corresponds to the incompressible Neo-Hookean model. For $\mathrm{O}_3$, the auxiliary variable is chosen as $y=\lambda^{-1/2}$, whereas for $\mathrm{O}_2$ and the MR model we use $y=\lambda^{-1}$. The resulting potential energies take the form
\begin{equation*}
\widetilde{\mathcal E}[u,y]
=
\mathcal E[u,y]
+
\frac{\beta}{2}
\int_{\Omega_0}
\left(
1-y^{p}\lambda
\right)^2
\,dX,
\end{equation*}
where $\mathcal E[u,y]$ is obtained from \eqref{eq:mardsen_1d} using the corresponding constitutive model, and
$p= 1$ for $\mathrm{O}_2$ and $\mathrm{MR}$ and $p=2$ for $\mathrm{O}_3$.  The current results are a proof of concept obtained via statevector simulation in Qiskit. A complexity analysis follows analogously from Section~5 of \cite{KO24}, which addresses hardware implementation challenges and the conditions for a potential exponential advantage. The problem parameters are chosen as $L=1$ mm, $\bar P=-2$ N/mm$^2$, $\bar u=0.3$ mm, $B(X)=10X$ N/mm$^3$ and $\beta = 10$. An $n=3$ qubit system ($N_q=8$) is employed, with the leftmost node prescribed through the Dirichlet boundary condition.
We consider the iterative correction formulation \eqref{eq:iterative_correction}, retaining terms up to fourth order in $u'$.
The displacement profiles and corresponding relative $L^2$ errors with respect to the analytical solutions are shown in Figures~\ref{fig:comparison_models_plot} and \ref{fig:comparison_models_table}, respectively. 

The proposed iterative correction procedure improves the accuracy of the penalty formulation for all constitutive models considered, with the largest improvement observed for the Mooney--Rivlin model. Successive correction problems become increasingly difficult to optimize as the solution approaches the analytical response. Consequently, the refinement process is terminated after five iterations, by which point substantial accuracy improvements have already been achieved.  While the present examples require only a single auxiliary variable, more general constitutive models may involve multiple fractional-power terms, leading to additional auxiliary variables and a correspondingly more expensive quantum implementation. Nevertheless, the same underlying construction remains applicable.

\section{Conclusion} \label{sec:conclusion}
The VQA framework proposed in \cite{KO24} was extended to a class of hyperelastic constitutive models whose nonlinearities can be represented through rational power-law expressions. The proposed methodology combines a penalty-based VQA for obtaining an approximate solution with a sequence of correction VQAs that iteratively improve the result. The approach was demonstrated using two variants of the Ogden model and the Mooney--Rivlin model. Future work will focus on developing more efficient treatments of complex nonlinearities, such as higher-order fractional powers appearing in \eqref{eq:general_nonlin}, while accounting for the associated implementation and complexity considerations. Another important direction is the extension of the framework to higher-dimensional problems, where additional theoretical and computational challenges are expected to arise. Leveraging new developments that point to efficient representation of the resulting systems of equations~\cite{PhiloOskay2026} is critical to evaluating multidimensional problems that involve unstructured grids that fully leverage the capabilities of the finite element method.

\section{Acknowledgements} \label{sec:acknowledgements}

We gratefully acknowledge the funding support from NSF Mechanics
of Materials and Structures Program (Award No: 2222404 and No: 2527249).



\bibliographystyle{IEEEtran}
\bibliography{references}

\end{document}

%% file: temp-definitions.tex
\usepackage{amsmath}
\usepackage{amsbsy}
\usepackage{amssymb}
\usepackage{amscd}
\usepackage{amsfonts}

\newcommand{\beq}{\begin{equation}}
\newcommand{\eeq}{\end{equation}}
\newcommand{\beqs}{\begin{eqnarray}}
\newcommand{\eeqs}{\end{eqnarray}}
\newcommand{\beql}{\begin{equation} \label}

\newcommand{\bbu}{\boldsymbol{u}}

\newcommand{\bbX}{\boldsymbol{X}}
\newcommand{\bbP}{\boldsymbol{P}}
\newcommand{\bbB}{\boldsymbol{B}}

\newcommand{\bbF}{\boldsymbol{F}}

\newcommand{\bbOmega}{\boldsymbol{\Omega}}


%% file: references.bib
@article{KO24,
  title         = {A Variational Quantum Algorithm for Nonlinear Finite Element Analysis of Hyperelastic Materials},
  author        = {Kouskiya, Uditnarayan and Oskay, Caglar},
  journal       = {arXiv preprint arXiv:2605.29181},
  year          = {2026},
  eprint        = {2605.29181},
  archivePrefix = {arXiv},
  primaryClass  = {quant-ph},
  doi           = {10.48550/arXiv.2605.29181}
}

@article{lubasch_zoo,
  title = {Variational quantum algorithms for nonlinear problems},
  author = {Lubasch, Michael and Joo, Jaewoo and Moinier, Pierre and Kiffner, Martin and Jaksch, Dieter},
  journal = {Phys. Rev. A},
  volume = {101},
  issue = {1},
  pages = {010301},
  numpages = {7},
  year = {2020},
  month = {1},
  publisher = {American Physical Society},
  doi = {10.1103/PhysRevA.101.010301},
  url = {}
}

@book{NocedalWright,
  author    = {Nocedal, J. and Wright, S. J.},
  title     = {Numerical Optimization},
  publisher = {Springer},
  edition   = {2},
  year      = {2006}
}

@book{marsden1994mathematical,
  title={Mathematical Foundations of Elasticity},
  author={Marsden, Jerrold E and Hughes, Thomas JR},
  year={1994},
  publisher={Dover Publications}
}

@article{sim_et_al,
author = {Sim, Sukin and Johnson, Peter D. and Aspuru-Guzik, Alán},
title = {Expressibility and Entangling Capability of Parameterized Quantum Circuits for Hybrid Quantum-Classical Algorithms},
journal = {Advanced Quantum Technologies},
volume = {2},
number = {12},
pages = {1900070},
doi = {https://doi.org/10.1002/qute.201900070},
url = {},
eprint = {https://advanced.onlinelibrary.wiley.com/doi/pdf/10.1002/qute.201900070},
year = {2019}
}

@article{Arora:2025a,
	author = {Arora, A. and Ward, B. M. and Oskay, C.},
	journal = {Finite Elements in Analysis and Design},
	pages = {104354},
	title = {An implementation of the finite element method in hybrid classical/quantum computers},
	volume = {248},
	year = {2025},
}

@book{Jog2015,
  author    = {C. S. Jog},
  title     = {Continuum Mechanics},
  series     = {Foundations and Applications of Mechanics},
  volume     = {1},
  publisher  = {Cambridge University Press},
  address    = {Cambridge},
  year       = {2015},
  edition    = {Revised},
  isbn       = {9781107091351}
}

@article{PhiloOskay2026,
  title         = {PHASE: Pauli Hierarchical Assembly on Subdivided Elements for Quantum-Compatible Operator Synthesis},
  author        = {Philo, Tillman and Oskay, Caglar},
  year          = {2026},
  eprint        = {2606.11478},
  archivePrefix = {arXiv},
  primaryClass  = {quant-ph},
  doi           = {10.48550/arXiv.2606.11478},
  url           = {https://arxiv.org/abs/2606.11478}
}
